\documentclass[prl,twocolumn,letterpaper, superscriptaddress, longbibliography ]{revtex4-1}

\usepackage{graphicx}
\usepackage{dcolumn}
\usepackage{bm}

\usepackage{graphicx}
\usepackage{bm}
\usepackage{amssymb}
\usepackage{amsmath}
\usepackage{dcolumn}
\usepackage{subfigure}
\usepackage{threeparttable}
\usepackage{multirow}
\usepackage[usenames,dvipsnames]{color}
\usepackage{wasysym}
\usepackage{txfonts}
\usepackage{mathtools}

\begin{document}

\preprint{ }

\title[title]{Self-hybridization and tunable magnon-magnon coupling in van der Waals synthetic magnets}
\author{Joseph Sklenar}
\email{Email: jnsklenar@wayne.edu}
\affiliation{Department of Physics and Astronomy, Wayne State University, Detroit MI 48202, USA}

\author{Wei Zhang}

\affiliation{Department of Physics, Oakland University, Rochester MI 48309, USA}


\date{\today}

\begin{abstract}
Van der Waals magnets are uniquely positioned at the intersection between two-dimensional materials, antiferromagnetic spintronics, and magnonics.  The interlayer exchange interaction in these materials enables antiferromagnetic resonances to be accessed at GHz frequencies.  Consequently, these layered antiferromagnets are intriguing materials out of which quantum hybrid magnonic devices can be fashioned. Here, we use both a modified macrospin model and micromagnetic simulations to demonstrate a comprehensive antiferromagnetic resonance spectra in van der Waals magnets near the ultrathin (monolayer) limit.  The number of optical and acoustic magnon modes, as well as the mode frequencies, are found to be exquisitely sensitive to the number of layers. We discover a self-hybridization effect where pairs of either optical or acoustic magnons are found to interact and self-hybridize through the dynamic exchange interaction.  This leads to characteristic avoided energy level crossings in the energy spectra. Through simulations, we show that by electrically controlling the damping of surface layers within heterostructures both the strength and number of avoided energy level crossings in the magnon spectra can be controlled.  
\end{abstract}


\maketitle
\section{Introduction}
The promise of antiferromagnetic (AFM) spintronics\cite{jungwirth2016antiferromagnetic, baltz2018antiferromagnetic, duine2018synthetic, siddiqui2020perspective,wz_mser2016} is usually ascribed to the fact that AFM materials possess ultrafast magnonic excitations.  These high frequency modes are considered technologically promising, suggesting that AFM memories that can switch at THz speeds\cite{vzelezny2014relativistic, cheng2016terahertz,roman_srep2017,junxue_nature2020,vaidya_science2020}.  In collinear antiferromagnets, there are two fundamental resonant modes differentiated by the relative phase difference of the precessing magnetization between the two magnetic sub-lattices\cite{keffer1952theory}.  These are the  optical and acoustic antiferromagnetic resonance (AFMR), or magnon modes if there is spatial variation in the phase of precession.  If we exclude magnetic anisotropy effects, the frequency of the optical modes is directly set by the exchange interaction across all length scales. This trait permits the excitation of spatially uniform modes which can span a wide range of frequencies (10 GHz - 2 THz) due to a corresponding wide range of antiferromagnetic exchange interactions\cite{macneill2019gigahertz, moriyama2019intrinsic, Li2020Nature,vaidya2020subterahertz}. 

Van der Waals (VdW) magnets are a new class of cleavable magnetic materials that remain magnetic in the ultrathin, two-dimensional limit\cite{huang2017layer,gong2017discovery}.  Some of the VdW magnets studied so far have an intralayer ferromagnetic interaction, and an interlayer antiferromagnetic interaction\cite{huang2017layer, mcguire2017magnetic, hu2020van}.  Thus, if an ultrathin sample has an even number of layers, the material behaves like an $A$-type antiferromagnet.  The interlayer antiferromagnetic exchange interaction is usually weaker than the intralayer ferromagnetic interaction.  This trait has enabled the observation of both optical and acoustic AFMR modes in CrCl$_3$ at GHz frequencies\cite{macneill2019gigahertz}.  In CrCl$_3$, tunneling magnetoresistance measurements have recently suggested that the interlayer exchange interaction can be enhanced in the small layer limit\cite{klein2019enhancement}.  The wide range of available magnon frequencies in VdW magnets implies a unique, tunable, application potential in these materials\cite{cortie2020two}.  However, due to atmospheric sensitivity of VdW magnets along with general difficulty in device fabrication, few experimental studies of magnetization dynamics in the ultrathin limit exist.  Through both Raman spectroscopy and time-resolved magneto-optical Kerr studies, there is early evidence that magnon frequencies in the two-dimensional limit can span a wide range of values up to many hundreds of GHz \cite{cenker2020direct, zhang2020gate}.  

The accessibility of optical magnon modes at lower frequencies (10-100 GHz) makes VdW magnets  intriguing from both a spintronics \textit{and} magnonics point-of-veiw.  Magnonics is a sub-field of magnetism which seeks to functionalize magnons, or spin waves, for the transmission and processing of information\cite{kruglyak2010magnonics,lenk2011building}.  Over the last few years, the field of magnonics has evolved to consider the potential of utilizing magnons within the quantum information sciences by building hybrid quantum systems e.g. a magnon-photon-qubit \cite{tabuchi2015coherent,lachance2019hybrid, lachance2020entanglement,li2020perspective} has been examined.  In order to build towards these grander aspirations, much work has been done involving the coherent coupling of magnons with microwave photons in both cavities\cite{bai2015spin} and, more recently, ``on-chip'' with lithographically patterned superconducting resonantors\cite{li2019strong,hou2019strong}.  Along similar lines, there are also emerging interests in hybridizing magnons with other magnons\cite{klingler2018spin, chen2018strong, qin2018exchange, li2020coherent, xiong2020MIT}.  In these studies of coherently coupled magnons, ferromagnetic materials are used.  The accessibility of long wavelength ferromagnetic magnons at GHz frequencies underlies this predilection to consider ferromagnetic materials.

The goal of this Article is to demonstrate that magnetization dynamics in VdW antiferromagnets provides a rich platform for studying the hybridization and coupling of magnons. Chromium trihalides such as CrI$_3$, CrCl$_3$, and CrBr$_3$\cite{kim2019evolution} belong to one of the best classes of materials for such future experiments.  To achieve this goal, we present both a macrospin and micromagnetic framework for examining both the optical and acoustic AFMR spectrum in CrCl$_3$, beyond the bilayer limit.  Although we focus our discussions on known material parameters for CrCl$_3$, our final set of conclusions is generalized to many similar materials, including synthetic antiferromagnetic materials\cite{duine2018synthetic}. 

Our results demonstrate that both the optical and acoustic magnon mode branches, frequencies, and interactions are sensitive to the number of layers present when near the monolayer limit. This result can be immediately taken into account for ongoing experiments attempting to measure magnetization dynamics in van der Waals materials near the ultrathin limit\cite{cenker2020direct, zhang2020gate}. In examining the AFMR spectra, we will pay particularly close attention to both the presence and interaction of various optical and acoustic modes that are available when more than two magnetic layers are present.  We will focus on the \textit{tetralayer}, or four-layered material, to emphasize these points. The tetralayer is the simplest example of a net antiferromagnetic layered material that is able to showcase the magnon-magnon hybridization effects  we are interested in.  The key to the self-hybridization effect we predict is that the magnonic excitations must be able to be localized on individual layers, while still being able to interact with one another.  No such localized modes exist for the bilayer i.e. only uniform antiferromagnetic resonance modes tend to be excited with uniform fields.  We will demonstrate that self-hybridized modes are unique compared to the recent studies which have focused on hybridizing optical and acoustic magnon modes.  Typically, this type of hybridization is enabled through a symmetry breaking external field\cite{macneill2019gigahertz, sud2020tunable} or through the dipolar interaction\cite{shiota2020tunable}.  In contrast, self-hybridized magnons are coupled through the dynamic exchange interaction.  This creates a circumstance  whereby, adjusting the damping on the surface layers, dramatic alterations in the magnon energy spectra take place, e.g., the strength and number of avoided energy level crossings can be directly controlled.  Because surface damping modifications can be achieved through all-electrical means, these results profer a new connection between two-dimensional materials with quantum magnonics, by offering viable ways such as to electrically control magnon-magnon interactions within VdW magnets.

\section{Micromagnetic and Macrospin Methodology}
We used two complementary approaches to study  magnetization dynamics in CrCl$_3$.  We simulated finite-sized CrCl$_3$ samples using Mumax3, a GPU enabled micromagnetic simulation software package\cite{vansteenkiste2014design}.  In most experiments, the shape of the exfoliated flakes are of irregular shape, providing a natural magnetic easy (dominant) axis due to shape anisotropy. Therefore, the geometry we chose for the CrCl$_3$ simulation was based on a standard example\cite{vansteenkiste2014design}, and is an ellipse of 160 nm long and 80 nm wide.  Because the in-plane magnetocrystalline anisotropy of CrCl$_3$ is negligible, the ellipse offers a natural ``easy'' axis through shape anisotropy, and we took advantage of the shape when setting single domain initial configurations.  An advantage of the micromagnetic approach is that it is easy to add additional layers, and in this work we consider up to six total layers.  The important material parameters for micromagnetic simulations are: the saturation magnetization ($M_s$) of a single layer (315 $A/m$), the exchange stiffness within a  layer (1.3 $\times 10^{-12}$ $J/m$), and the antiferromagnetic exchange coupling between layers which is 0.32$\%$ of the ferromagnetic exchange coupling.  These material properties are taken from the known low-temperature values in bulk CrCl$_3$\cite{mcguire2017magnetic, macneill2019gigahertz}.  In all simulations we started with a zero-field configuration, where the magnetization of each layer is staggered along the $\pm x$-direction which is aligned along the long axis of the ellipse.  We then considered magnetostatic configurations, and the resulting dynamics of these configurations when a field was applied perpendicular, along the $y$-axis. 

The macrospin approach has already been used to study bilayers, and in this work we extended the macrospin model to examine the more complicated tetralayer geometry.  In the macrospin formulation, each layer is modeled as a single rigid magnetic moment which is antiferromagnetically coupled to adjacent macrospins.  The equations of motion can be written down using the Landau-Lifshitz-Gilbert (LLG) equation, and there are $N$ coupled equations, where $N$ is the number of layers.  So, for a bilayer system the macrospin equations of motion are:
\begin{equation}
  \begin{gathered}
    \frac{d\mathbf{\hat{m}_A}}{dt} = -\mu_0\gamma \mathbf{\hat{m}_A} \times [H_0\hat{y} - H_E\mathbf{\hat{m}_B} -M_s(\mathbf{\hat{m}_A} \cdot \hat{z})\hat{z}],\\
    \frac{d\mathbf{\hat{m}_B}}{dt} = -\mu_0\gamma \mathbf{\hat{m}_B} \times [H_0\hat{y} - H_E\mathbf{\hat{m}_A} -M_s(\mathbf{\hat{m}_B} \cdot \hat{z})\hat{z}].
  \end{gathered}
  \label{eq:bilayer_eom}
\end{equation}
Similarly, for a tetralayer system the equations of motion are:
\begin{equation}
  \begin{gathered}
    \frac{d\mathbf{\hat{m}_A}}{dt} = -\mu_0\gamma \mathbf{\hat{m}_A} \times [H_0\hat{y} - H_E\mathbf{\hat{m}_B} -M_s(\mathbf{\hat{m}_A} \cdot \hat{z})\hat{z}],\\
    \frac{d\mathbf{\hat{m}_B}}{dt} = -\mu_0\gamma \mathbf{\hat{m}_B} \times [H_0\hat{y} - H_E\mathbf{\hat{m}_A} - H_E\mathbf{\hat{m}_C} - M_s(\mathbf{\hat{m}_B} \cdot \hat{z})\hat{z}],\\
    \frac{d\mathbf{\hat{m}_C}}{dt} = -\mu_0\gamma \mathbf{\hat{m}_C} \times [H_0\hat{y} - H_E\mathbf{\hat{m}_B} - H_E\mathbf{\hat{m}_D} - M_s(\mathbf{\hat{m}_C} \cdot \hat{z})\hat{z}],\\
    \frac{d\mathbf{\hat{m}_D}}{dt} = -\mu_0\gamma \mathbf{\hat{m}_D} \times [H_0\hat{y} - H_E\mathbf{\hat{m}_C} -M_s(\mathbf{\hat{m}_D} \cdot \hat{z})\hat{z}].
  \end{gathered}
   \label{eq:tetralayer_eom}
\end{equation}
In the above equations, the unit vectors (e.g. $\mathbf{\hat{m}_A}$) refer to the direction that a given macrospin is pointing, and the subscript references the individual layer.  ${H}_{0}$ is the externally applied magnetic field.  The interlayer exchange field strength is given by $H_E$, and the saturation magnetization is denoted by $M_s$.  The last term in each equation of motion is the demagnetization field from thin film shape anisotropy.  It should be noted that for the tetralayer, layers \textbf{B} and \textbf{C} experience two separate exchange fields owing to the fact that the interior layers are coupled to two layers each, while the surface layers are only coupled to one layer.  

\subsection{Static Magnetization Configurations}

\begin{figure*}
\includegraphics[scale = .24]{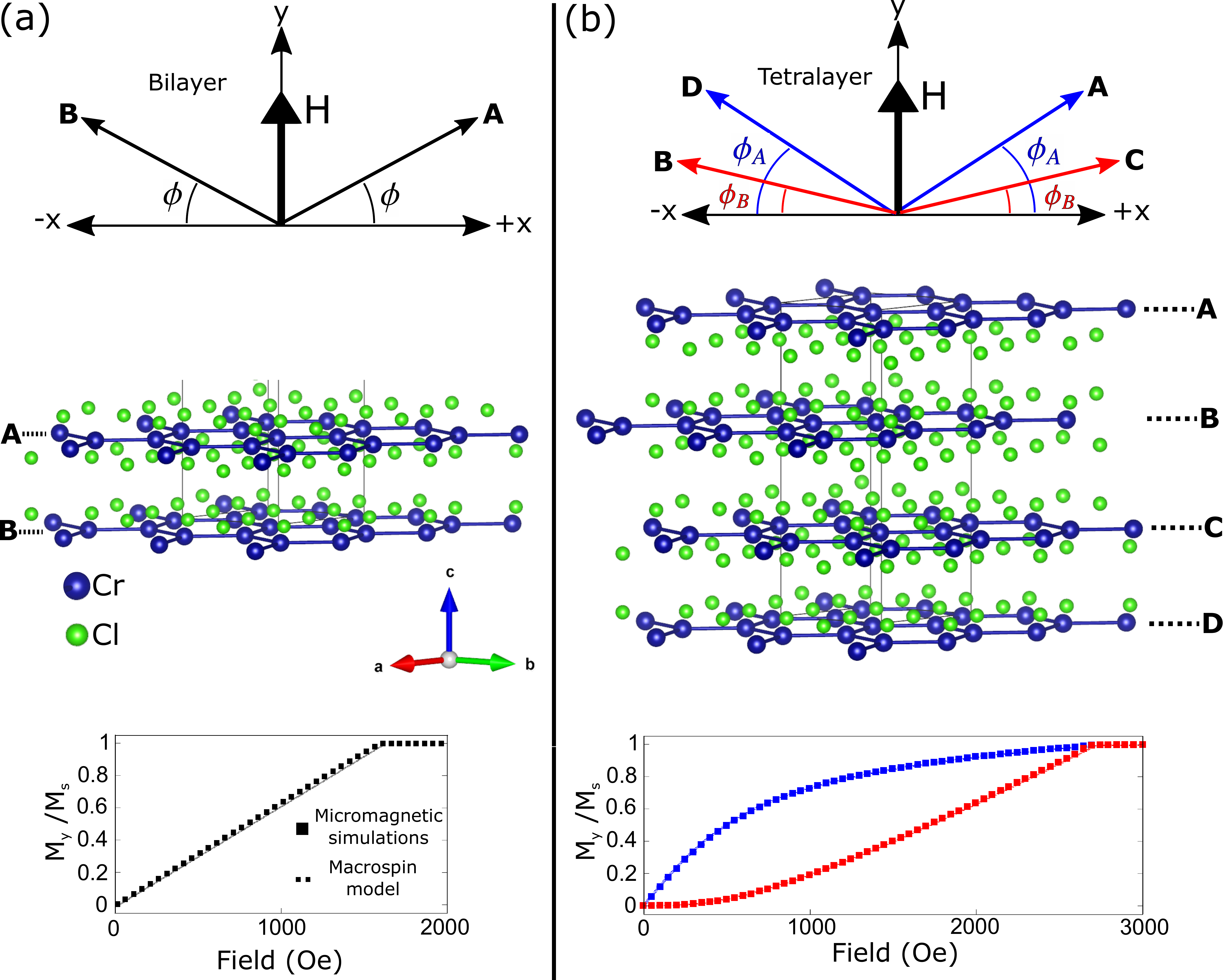}
\caption{(a) The static equilibrium configurations of a CrCl$_3$ bilayer are shown.  In the macrospin model, the static equilibrium orientation of $\hat{m}_A$ and $\hat{m}_B$ can be described by a single angle $\phi$ which depends on the external field, as illustrated.  The layered CrCl$_3$ structure corresponding to the bilayer is also drawn.  At the bottom of (a), the normalized magnetization projection along the external field direction is shown for the bilayer; results are plotted from both the macrospin model and micromagnetic simulations.  (b) The static equilibrium configurations of a CrCl$_3$ tetralayer are shown.  For the tetralayer, two angles, $\phi_A$ and $\phi_B$, are needed to describe the static equilibrium orientation of the surface and interior layers respectively as a function of field.  The layered CrCl$_3$ structure corresponding to the tetralayer is drawn below the macrospin image. At the bottom of (b), the normalized magnetization projection along the external field direction is shown for the tetralayer; results are plotted from both the macrospin model and micromagnetic simulations.  Note that the blue squares and lines represent the surface layers while the red counterparts represent the interior layers.  In both panels, the CrCl$_3$ illustrations were generated using VESTA \cite{momma2011vesta}.}
\label{fig:static}
\end{figure*}  

Before considering the layer-dependent dynamic properties of CrCl$_3$, we obtained static equilibrium magnetization configurations as a function of the external magnetic field.  In all discussions that follow, we set the equilibrium positions of the macrospins along the $\pm x$-axis in the absence of a magnetic field.  Similarly, we aligned the easy axis of the micromagnetic ellipse along the $x$-axis so that the micromagnetic configuration without a field had the magnetization of each layer alternating along the easy axis.  When an external field is applied along the $y$-direction, a new equilibrium configuration must then be obtained for both the macrospin model and micromagnetic simulations.

In the macrospin model, the equilibrium orientation was obtained by setting the net torque on each individual macrospin to zero.  For bilayers, only one angle ($\phi$) is needed to describe the equilibrium orientation of the macrospins as illustrated in Fig.~\ref{fig:static} (a).  It was previously shown that the equilibrium angle as a function of the external field can be obtained by solving: $\sin{\phi} = H/2H_E$\cite{macneill2019gigahertz}. For tetralayers, two angles ($\phi_A$ $\&$ $\phi_B$) are needed to describe the equilibrium configuration as seen in Fig.~\ref{fig:static} (b).  $\phi_A$ refers to the tilting of the surface layers away from the $x$-axis, while $\phi_B$ describes the tilting of the two interior layers.  As a function of the external field, $\phi_A$ and $\phi_B$ were obtained by enforcing the equilibrium condition on Eq.~\ref{eq:tetralayer_eom}; i.e. setting the net torques on all macrospins equal to zero.  After enforcing the static equilibrium condition, the following pair of equations can be generated, and must be solved numerically to find the equilibrium angles as a function of field:
\begin{equation}
  \begin{gathered}
    H_0\cos{\phi_A} - H_E\sin{(\phi_A + \phi_B)} = 0,\\
    H_0\cos{\phi_B} - H_E[\sin{(\phi_A + \phi_B)}+\sin{(2\phi_B)}] = 0.\\
  \end{gathered}
  \label{eq:static_eq}
\end{equation}

When using micromagnetic simulations, the equilibrium orientations were obtained by applying a field and using an energy minimization algorithm built into Mumax3 to find the resulting static configuration.  This strategy can be used for simulations containing any number of layers.  In Fig.~\ref{fig:static} (a) and (b), the CrCl$_3$ layered crystal structure is shown for a bilayer and tetralayer.  In micromagnetic simulations, we use micromagnetic cells two layers and four layers thick to represent the CrCl$_3$ bilayer and tetralayer structures.  We use a micromagnetic cell, 0.6 nm thick, to mimic the thickness of an individual layer of CrCl$_3$\cite{cai2019atomically}.   For clarity and reference, we have labeled the individual layers, \textbf{A,B,C,D}, consistent with the macrospin labeling.

In Fig.~\ref{fig:static} (a) and (b), we summarize the static equilibrium orientations as a function of the external field for the bilayer and tetralayer respectively.  We plot the projection of the normalized magnetization along the $y$-direction as a function of field.  For micromagnetic simulations this is done by averaging the magnetization on each individual layer.  Square symbols represent results from micromagnetic simulations while the dashed lines represent macrospin results obtained by numerically solving Eq.~\ref{eq:static_eq}.  For the tetralayer, blue color represents the surface moments (layers \textbf{A,D}) and red color represents the interior moments (layers \textbf{B,C}).  The static equilibrium orientations of individual macrospins, and the averaged micromagnetic magnetization are in close agreement between both models.

\subsection{Dynamic Magnetization Calculations}
We now discuss how the antiferromagnetic resonance modes can be calculated using both the macrospin model and micromagnetic simulations.  For bilayers, the macrospin treatment involves linearization of the equations of motion (Eq.~\ref{eq:bilayer_eom}), and these have been solved elsewhere\cite{macneill2019gigahertz}. Here, we state the previous result for completeness: 
\begin{equation}
    \omega_{2,O} = \mu_0\gamma\sqrt{2H_EM_s\left(1-\frac{H_0^2}{4H^2_E}\right)},
    \label{eq:bi_O}
\end{equation}
\begin{equation}
    \omega_{2,A} = \mu_0\gamma\sqrt{2H_E\left(2H_E + M_s\right)}\frac{H_0}{2H_E}.
    \label{eq:bi_A}
\end{equation}
In the above equations $\omega_{2,O}$ refers to the optical while $\omega_{2,A}$ corresponds to the acoustic antiferromagnetic resonance modes, respectively.

\begin{figure}
\includegraphics[scale = .145]{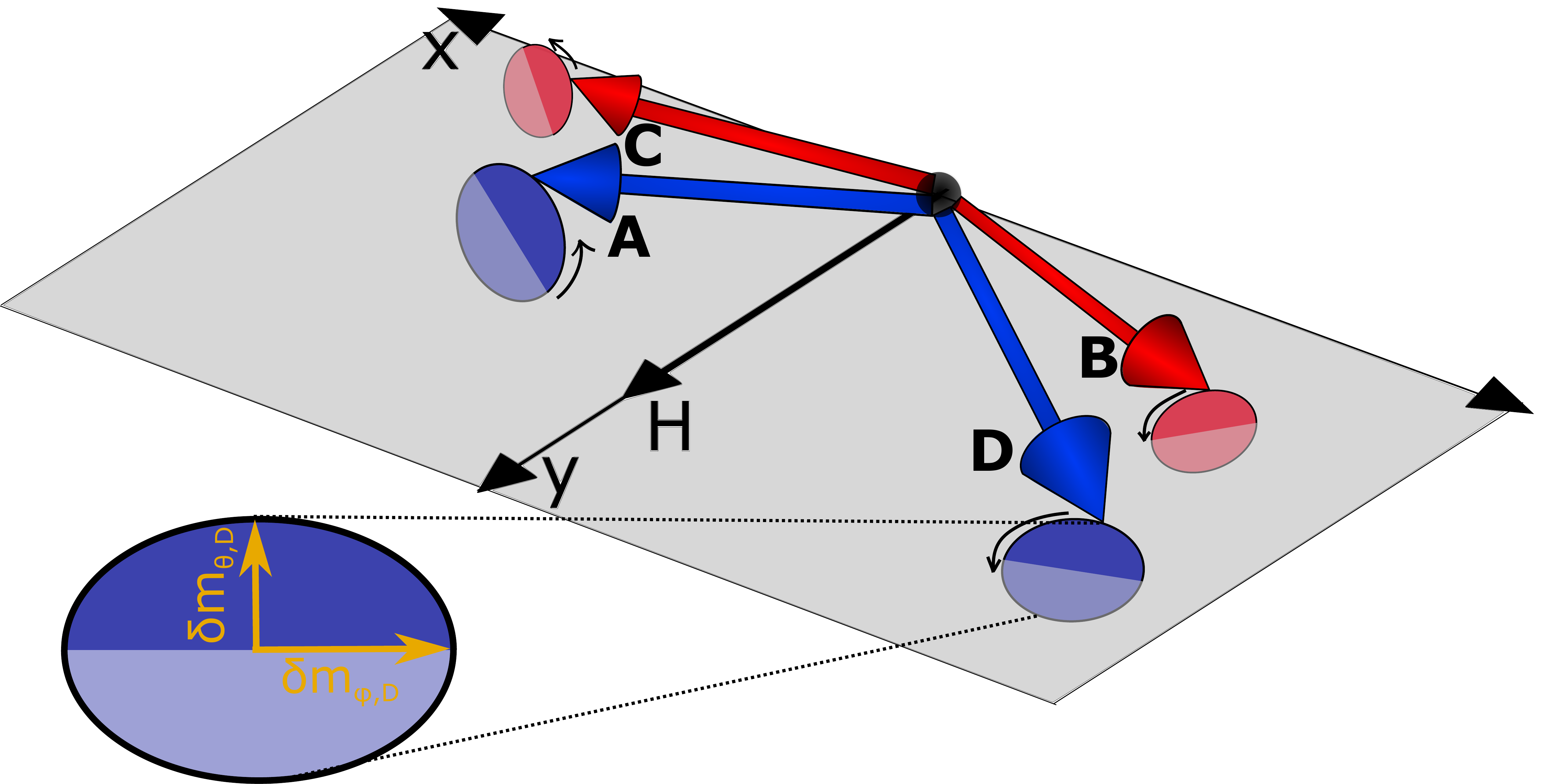}
\caption{The magnetization dynamics of the tetralayer, using the macrospin model, are illustrated. The dynamics of a given macrospin can be described with two amplitudes.  We illustrated the dynamics for layer \textbf{D}, where $\delta m_{\phi,D}$ and $\delta m_{\theta,D}$ correspond to the in-plane and out-of-plane dynamic amplitudes.  }
\label{fig:macrodyna}
\end{figure}

To solve for the antiferromagnetic resonances of the tetralayer, we linearized  Eqs.~\ref{eq:tetralayer_eom} using rotated coordinate systems, ($x'y'z'$), where the static equilibrium orientation of a given macrospin lies along the $x'$-direction.  In this new type of coordinate system $\mathbf{\hat{m}_A}$ can be written as:  $\mathbf{\hat{m}_A} = (1, e^{i\omega t}\delta m_{\phi,A},e^{i\omega t}\delta m_{\theta,A}) = (1,\mathbf{\delta m_A}(t))$.  The dynamic amplitudes, $\delta m_{\phi,A}$ and $\delta m_{\theta,A}$, describe the elliptical path traced out during the precession of the macrospin about the direction of the static equilibrium direction.  These amplitudes are illustrated in Fig.~\ref{fig:macrodyna}.  The system of equations describing the tetralayer can be linearized with respect to the dynamic amplitudes, and the linearized version of the first two lines of Eqs.~\ref{eq:tetralayer_eom} are:
\begin{equation}
  \begin{gathered}
    i \omega\mathbf{\delta m_A} = \mu_0\gamma \mathbf{\hat{m}_A} \times [H_{A,eq}\mathbf{\delta m_A}  + H_E\mathbf{\delta m_B}  +M_s(\mathbf{\delta m_A} \cdot \hat{z})\hat{z}]\\
    i \omega\mathbf{\delta m_B} = \mu_0\gamma \mathbf{\hat{m}_B} \times [H_{B,eq}\mathbf{\delta m_B}  + H_E\mathbf{\delta m_A} + H_E\mathbf{\delta m_C} \\  +M_s(\mathbf{\delta m_B} \cdot \hat{z})\hat{z}]
  \end{gathered}
  \label{eq:linear_tetra}
\end{equation}
Here, $H_{A,eq}$ and $H_{B,eq}$ are magnitudes of the net effective field that the surface and interior layers experience in static equilibrium.  These equilibrium fields are found numerically after $\phi_A$ and $\phi_B$ are obtained as a function of $H_0$.  For the tetralayer, we exploit the symmetries between the surface layers and the interior layers so that it is not necessary to linearize all four equations to obtain the optical and acoustic eigenmodes.  This is done by substituting $\mathbf{\delta m_C} = \pm\mathbf{\delta m_B}$, into Eqs.~\ref{eq:linear_tetra} in order to solve for the acoustic and optical field-frequency relationships respectively.  The equations describing the optical and acoustic eignenmodes are then:
\begin{widetext}
\begin{equation}
\begin{multlined} 
    \omega^4 - (\mu_0 \gamma \omega)^2[(-H_{B,eq}\pm H_E\cos{2\phi_B})(-H_{B,eq}\pm H_E-M_s) - (2\cos{(\phi_A+\phi_B)}H_E^2-H_{A,eq}^2 - H_{A,eq}M_s)] - \\ (\mu_0 \gamma)^4 \bigl[(H_{B,eq} \pm \cos{2\phi_B}H_E)[(H_{A,eq}^2+H_{A,eq}M_s)(-H_{B,eq}\pm H_E - M_s) + H_{A,eq}H_E^2] + \\  
    \cos{(\phi_A+\phi_B)}[(H_{B,eq}\pm H_E -M_s)(\cos{(\phi_A+\phi_B)}H_{A,eq}H_E +\cos{(\phi_A+\phi_B)}H_EM_s) + \cos{(\phi_A+\phi_B)}H_E^3] \bigr] = 0.
\end{multlined}
\label{eq:tetra}
\end{equation}
\end{widetext}
To obtain the field-frequency relationships of the optical and acoustic resonances, the roots of Eq.~\ref{eq:tetra} must be solved numerically.  In Eq.~\ref{eq:tetra}, the $\pm$ symbol refers to the differentiation between the optical modes ($+$) and the acoustic modes ($-$).

We now summarize how the dynamic modes were calculated from micromagnetic simulations.  First, the state of a given multilayer sample was initialized so that the magnetization on every layer is uniform and alternating on every layer along the $\pm x$-direction.  An external magnetic field was applied along the $y$-direction and the multilayer is allowed to relax into the static equilibrium configuration.  Magnetization dynamics are excited by either applying an external field pulse either along the $x$-direction or $y$-direction.   We used a 100 ps field pulse with an amplitude of 7 Oe to excite the dynamics.  After the pulse was applied, the micromagnetic moments were time-evolved according to the LLG equation for 10 ns.   To obtain a ``global'' picture of the resonant modes, we averaged the magnetization over every micromagnetic cell in the system at every time step.  This allowed us to generate a time series of the average magnetization: $\mathbf{M}_{avg}(t)$ = $(M^x_{avg}(t), M^y_{avg}(t), M^z_{avg}(t))$.  If the system was excited with a pulse along the $y$-direction, optical magnon modes were excited, and the average magnetization of the system has a temporally oscillating magnetic dipole moment along the $y$-direction.  By taking a fast Fourier transform (FFT) at every applied field of the $y$-component of the average magnetization, $M^y_{avg}(t)$, we are able to calculate the optical mode spectrum.  For acoustic modes to be excited, the field pulse is applied along the $x$-direction of the sample.  Unlike optical modes, acoustic modes will tend to have a non-vanishing average net moment along the $z$-direction.  Therefore, to calculate the acoustic mode spectrum, we took a FFT of $M^z_{avg}(t)$ at every field value.  

\begin{figure}
\includegraphics[scale = .3]{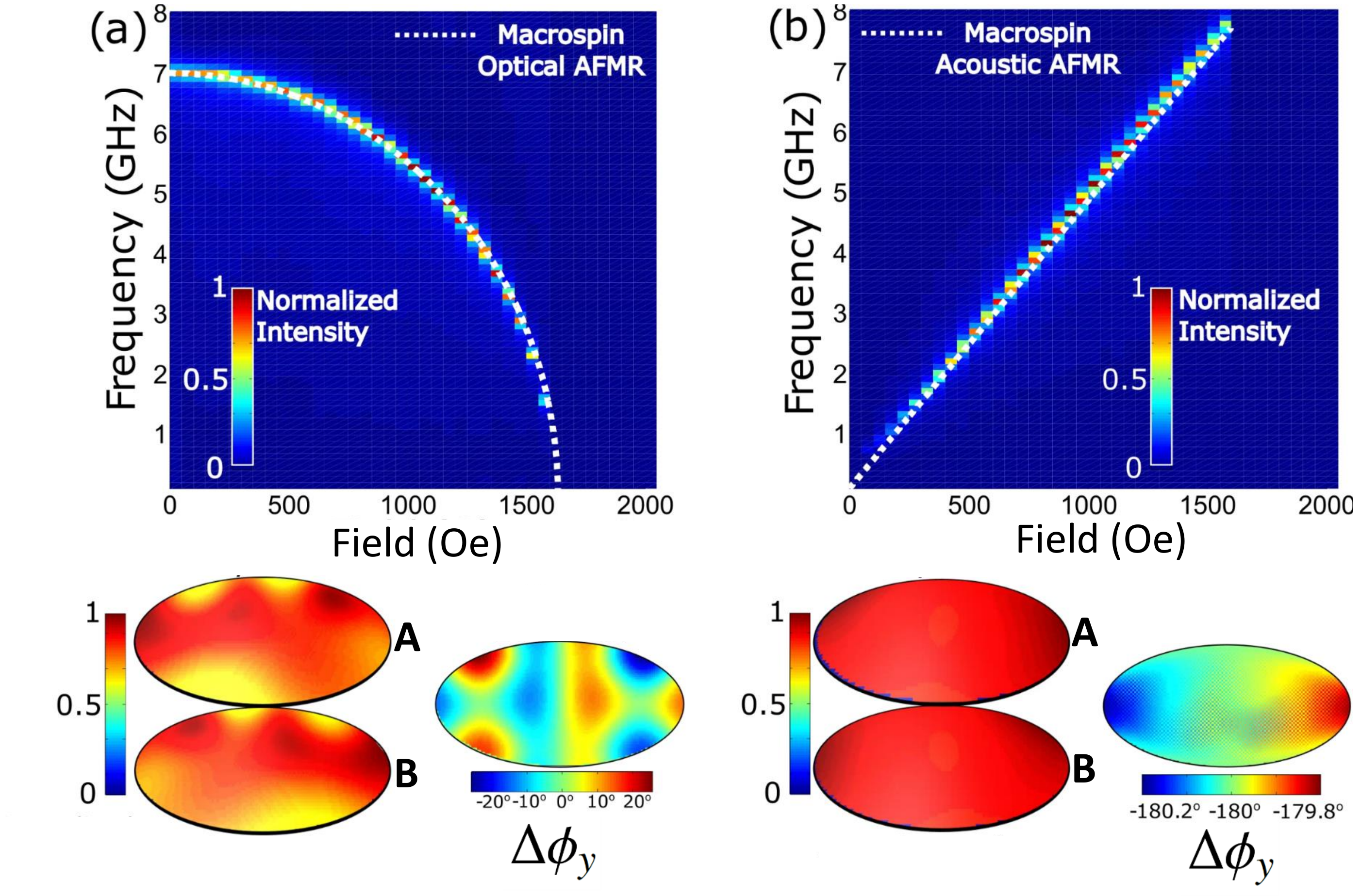}
\caption{(a) The global  AFMR spectra for the optical mode in the bilayer is shown for both the micromagnetic simulations (colormap) with the macrospin model results overlaid (white-dashed line).  The spatial resolution of the optical mode at an applied field of 500 Oe is also shown.  There is some spatial inhomogeneity of the mode within a given layer, but when comparing both layers it appears to be a quasi-uniform mode.  The color scale for the amplitude is normalized to the largest FFT amplitude in of a micromagnetic cell in either layer.  Finally, in (a), the phase difference, $\Delta\phi_y$, between layer \textbf{A} and \textbf{B} is calculated.  Although there is some inhomogenetity, the phase map is clearly centered around $\Delta\phi_y$ = $0^\circ$, which indicates an optical character of the mode.  (b)  The global AFMR spectra for the acoustic mode is shown for both the micromagnetic simulations and the macrospin model.  The spatial resolution of the acoustic mode at 500 Oe is also shown, and it is clearly a uniform mode.  Finally, the spatial resolution of $\Delta\phi_y$ is calculated.  The phase map is very uniform and centered around $\Delta\phi_y$ = $180^\circ$, indicating a uniform acoustic AFMR.  Note:  In (a) and (b) the micromagnetic color scale is normalized to the largest amplitude of the FFT across all field values.  \textit{We will use this normalized color scale throughout the entire document}.     }
\label{fig:bilayer}
\end{figure}  

\begin{figure*}
\includegraphics[scale = .52]{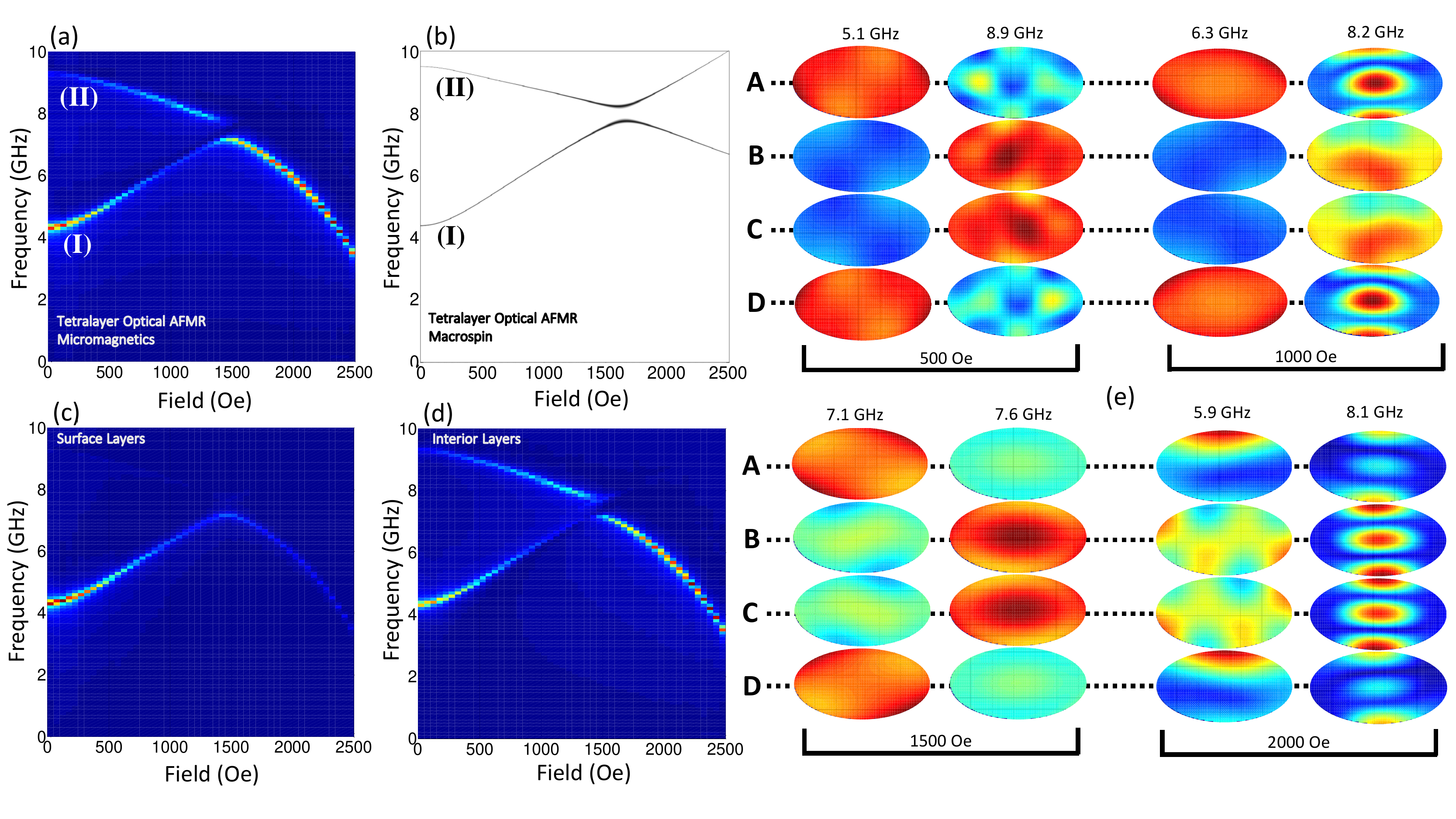}
\caption{(a) The optical mode spectra for the magnon modes are obtained from micromagnetic simulations by taking a FFT of the $y$-component of the globally averaged magnetization dynamics.  (b) The macrospin optical mode eigenfrequencies of the tetralayer are obtained by numerically solving Eq.~\ref{eq:tetra}.  (c) and (d) The micromagnetic mode spectra for the surface and interior layers are respectively shown after taking a FFT of the averaged magnetization dynamics on a layer-by-layer level.  (e) Spatial resolution of the micromagnetic optical magnon modes are shown for a selection of applied magnetic fields (500, 1000, 1500, and 2000 Oe).  For every field, the two mode frequencies are labeled.  We also denote the individual layers where $\mathbf{A}$ and $\mathbf{D}$ are the surface layers, and $\mathbf{B}$ and $\mathbf{C}$ are the interior layers.  For plots (a), (c), and (d), the color scale on every individual plot is normalized to the largest amplitude in the field-frequency spectra.  In (e), when spatially mapping an individual mode, we  have normalized the color scale to the largest amplitude on any of the four layers for a given field-frequency pairing.  }
\label{fig:tetra_optical}
\end{figure*}

It is often useful to adopt the above procedure on a layer-by-layer basis.  For the tetralayer, we extracted four different times series corresponding to the average magnetization of each layer, and took the FFT of each.  A layer-by-layer analysis provides information as to whether or not a mode is localized to certain layers or is uniform across all layers.  When hybridization between two magnon modes occurs, this analysis is a necessary first step towards the identification of the two modes that are interacting.   We are able to go one step further and spatially resolve the mode intensity on every layer as well the phase difference between layers.  To do this, we needed to take the FFT of the time series of each individual micromagnetic cell.  Because this process is more intensive from a data generation and processing point-of-view, we only performed this analysis at particular fields of interest.   

\subsection{Phase Resolution}
One way to confirm the optical or acoustic character of the magnon modes excited in the micromagnetic simulations is to spatially map out the phase of the dynamic component of the magnetization, $\Delta\phi_y$.  We define $\Delta\phi_y$ as the relative phase angle between micromagnetic cells in adjacent layers:  
\begin{equation}
  \Delta\phi_{y} = \arctan{\left(\frac{\Im[m^y_i(\omega)]}{\Re[m^y_i(\omega)]}\right)} - \arctan{\left(\frac{\Im[m^y_j(\omega)]}{\Re[m^y_j(\omega)]}\right)}  .
\end{equation}
Here, $m^y_{i}(\omega)$ refers to the FFT of $m^y_{i}(t)$, which is the time series of the $y$-component of the magnetization. The subscripts, $i$ and $j$, refer to the layers under consideration such as $\mathbf{A}$, $\mathbf{B}$, $\mathbf{C}$, or $\mathbf{D}$.  Typically we will calculate $\Delta\phi_y$ for adjacent layers e.g. $\mathbf{A}$ $\&$ $\mathbf{B}$ -or- $\mathbf{B}$ $\&$ $\mathbf{C}$.  By spatially mapping $\Delta\phi_y$ at field-frequency combinations which correspond to the excitation of a given magnon mode, we can check how homogeneous the phase differences between layers are.  More importantly, we are able to  obtain the average phase difference between layers.  This latter point allows unambiguous identification of acoustic or optical excitations between layers.  Since the external field in our work is applied along the $y$-direction, an optical excitation corresponds to $\Delta\phi_y$ = 0$^\circ$ while an acoustic excitation corresponds to $\Delta\phi_y$ = 180$^\circ$.

\section{Layer-Dependent Magnetization Dynamics}
In this Section we discuss the evolution of both the optical and acoustic antiferromagnetic resonance modes as the number of layers in CrCl$_3$ is increased.  We compare and contrast both the macrospin model and micromagnetic simulations. 

\subsection{Bilayer}

First, we briefly discuss the magnetization dynamics in a bilayer using both the macrospin analysis and micromagnetic simulations.  Bilayer macrospin models are frequently used in the literature\cite{lee2015macroscopic,liu2019observation,sud2020tunable}, and have recently been used to model very thick CrCl$_3$ platelets\cite{macneill2019gigahertz}. In the very thick limit, the effective field strength every layer experiences is nearly identical and a bilayer with identical layers shares this trait.  Both micromagnetic and macrospin results are shown in Fig.~\ref{fig:bilayer}.  Panels (a) and (b) plot a normalized intensity map of the global mode spectrum for the optical and acoustic mode respectively.  There is only one observed optical and acoustic AFMR branch in the bilayer.  The macrospin field-frequency behaviors, given by Eqs.~\ref{eq:bi_O} and ~\ref{eq:bi_A}, are overlaid on the intensity maps as dashed white lines.  There is clear quantitative agreement between both the micromagnetic and macrospin treatment of the bilayer.   As seen in Fig.~\ref{fig:bilayer} (a) and (b), spatially resolved maps of the mode intensity at 500 Oe indicate that each mode is quasi-uniform in both layers.  The acoustic mode, relative to the optical mode, is slightly more uniform.  We also spatially map the phase difference between layers for both the optical and acoustic mode in (a) and (b).  We find that the optical mode has a phase difference centered around $\Delta \phi_y$ = 0$^\circ$ while the acoustic mode is centered around $\Delta \phi_y$ = 180$^\circ$, which confirms the expected behavior. 
We are also able to conclude that even though spatial inhomogeneities are present in a more realistic micromagnetic object, these inhomogeneities do not lead to notable deviations in the field-frequency behavior compared with the macrospin results.



\subsection{Tetralayer Optical Modes}

We begin our discussion of the tetralayer by considering the optical mode spectrum.  In Fig.~\ref{fig:tetra_optical} (a) and (b) we plot the field-frequency behavior of the optical modes obtained from micromagnetic simulations and the macrospin model.  Unlike the bilayer, two optical mode branches are present.  A high frequency branch which decreases in frequency as a function of field, and a lower frequency branch which increases in frequency as the field is increased.  Both approaches to modeling the tetralayer clearly show an avoided energy crossing between the two branches; this suggests hybridization between the two optical modes.  The mode hybridization is enabled by the exchange coupling between the dynamic components of the magnetization.  This was verified in the micromagnetic simulations by, unphysically, disabling the exchange interaction related to the magnetization dynamics and not the static magnetization within the simulation. Based upon how the frequencies of the two modes change as the field is increased, a natural conjecture is that at low fields,  the low-frequency mode resides on the surfaces of the tetralayer while the high-frequency mode resides on the interior layers.  

\begin{figure}
\includegraphics[scale = .25]{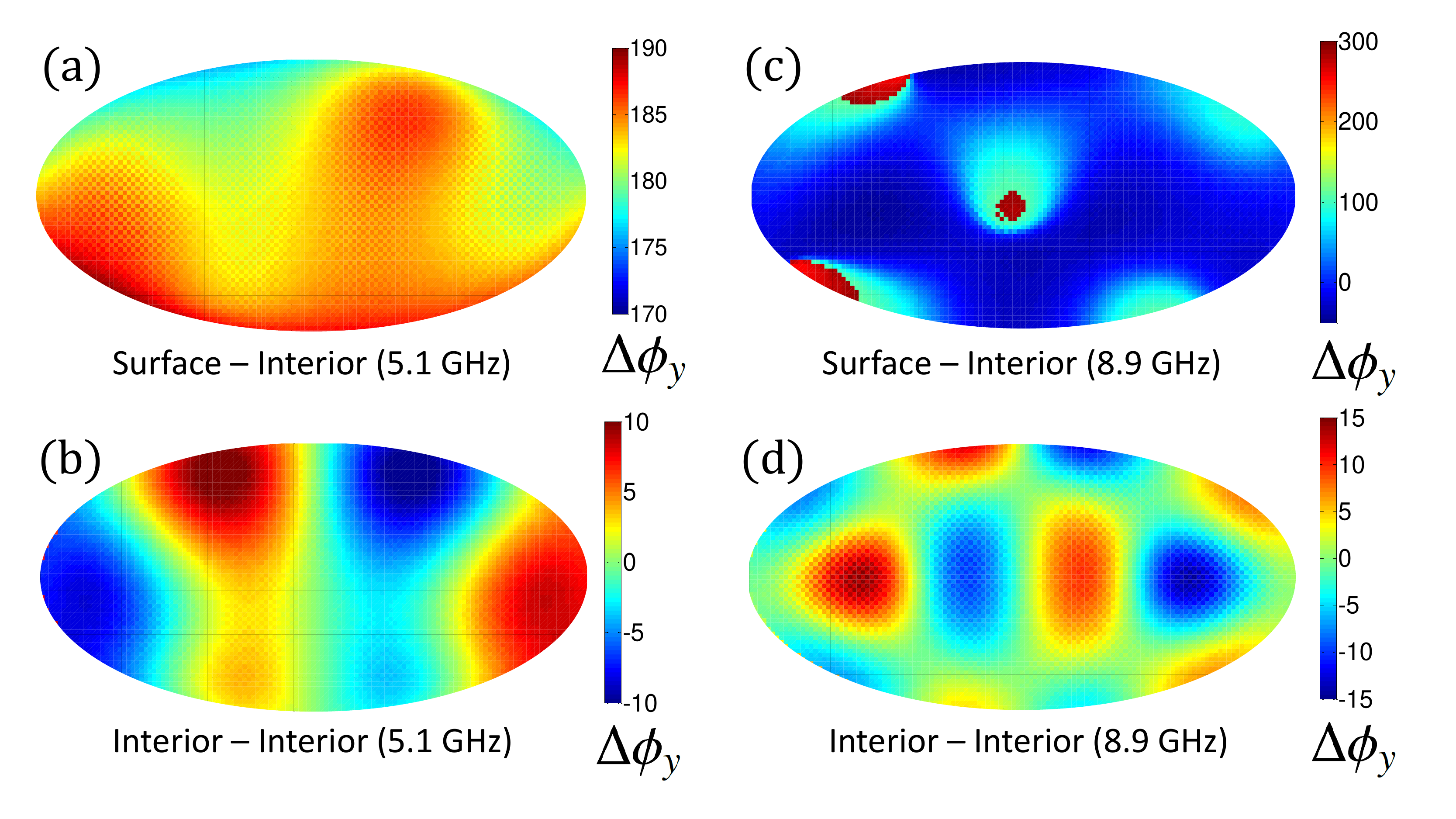}
\caption{(a) The phase difference, $\Delta \phi_y$, between a surface and interior layer of a tetralayer is spatially mapped at 5.1 GHz and 500 Oe.  (b) At the same field-frequency pair of (a), the phase difference between the two interior layers is spatially mapped.  The phase difference between the surface layers, where the magnetization is most strongly excited, and the interior layer is  $\Delta \phi_y$ = 0$^\circ$.  The phase difference between the two interior layers, where the magnetization is more weakly excited is centered around $\Delta \phi_y$ = 0$^\circ$.  This implies that the phase difference between the two surface layers is 0$^\circ$, or that the low frequency branch is an optical mode residing on the surfaces.   (c) The phase difference between a surface and interior layer of the mode at 8.9 GHz at 500 Oe is mapped.  (d)  For the same field-frequency pair in (c), the phase difference between the two interior layers is spatially mapped.  In (c), $\Delta \phi_y$ tends to be centered around 0$^\circ$. In (d) $\Delta \phi_y$, between the two interior layers where the dynamics are strongly excited, is centered around 180$^\circ$.  Thus the higher frequency mode is a more ``pure'' optical excitation that tends to be localized to the interior layers.   }
\label{fig:tetra_optical_phase}
\end{figure}  

The above assertion, on where the modes are localized, can be qualitatively explained by considering the zero-field intercept of the low-frequency branch (~4.2 GHz) compared with the high-frequency branch (~9.5 GHz).  For convenience, we have labeled the low-frequency branch as \textbf{I} and the high-frequency branch as \textbf{II} in Fig.~\ref{fig:tetra_optical} (a) and (b).    An optical mode localized on the surfaces has a lower energy because the magnetization dynamics are localized on two layers that are not exchange coupled to one another.  However, dynamics on the surface layers still incur an exchange energy cost when the magnetization in layer $\mathbf{A}$ moves away from being anti-parallel to the relatively static magnetization in layer $\mathbf{B}$.   On the other hand, when an optical mode is excited within the interior layers, the energy is slightly greater than twice that of the surface modes.  This is primarily because the static exchange field experienced in the interior layers is twice that of what a surface layer feels at zero field.  There is also exchange energy incurred from the dynamics; as the magnetization dynamics within the two adjacent interior layers is activated, the orientation between the magnetization in both layers (on average) is less anti-parallel than in the static configuration.

\begin{figure*}
\includegraphics[scale = .52]{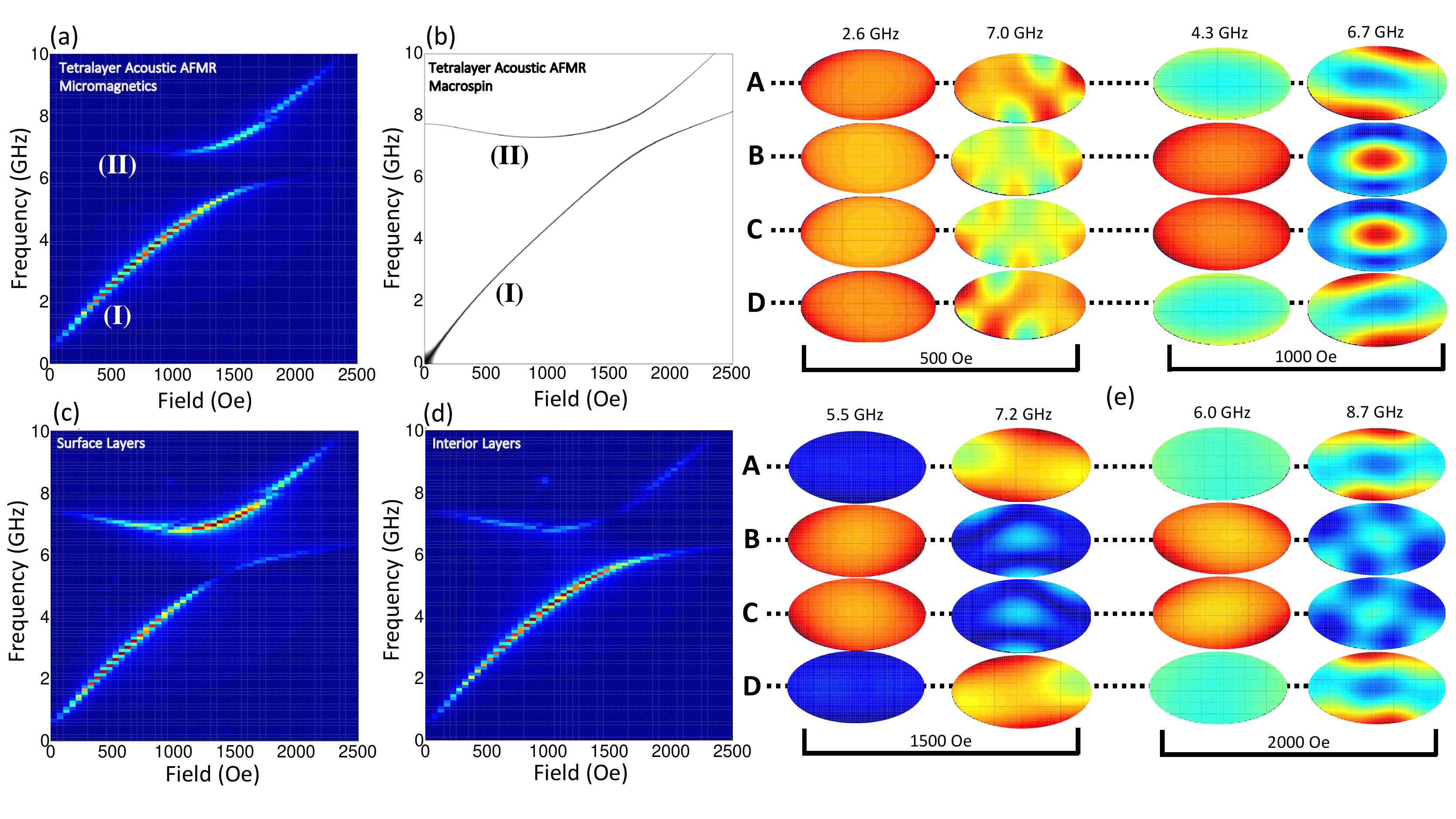}
\caption{(a) The acoustic mode spectra for the magnon modes are obtained from micromagnetic simulations by taking a FFT of the $z$-component of the globally averaged magnetization dynamics.  (b) The macrospin acoustic mode eigenfrequencies of the tetralayer are obtained by numerically solving Eq.~\ref{eq:tetra}.  (c) and (d) The micromagnetic mode spectra for the surface and interior layers are respectively shown after taking a FFT of the averaged magnetization dynamics on a layer-by-layer level.  (e) Spatial resolution of the micromagnetic acoustic magnon modes are shown for a selection of applied magnetic fields (500, 1000, 1500, and 2000 Oe).  For every field, the two mode frequencies are labeled.  We also denote the individual layers where $\mathbf{A}$ and $\mathbf{D}$ are the surface layers, and $\mathbf{B}$ and $\mathbf{C}$ are the interior layers.  For plots (a), (c), and (d), the color scale on every individual plot is normalized to the largest amplitude in the field-frequency spectra.  In (e), when spatially mapping an individual mode, we  have normalized the color scale to the largest amplitude on any of the four layers for a given field-frequency pairing.  }
\label{fig:tetra_acoustic}
\end{figure*}

We can verify and visualize the key points made in the preceding paragraph using micromagnetic simulations by: 1) Examining the field-frequency behavior of the power spectrum on a layer-by-layer basis and; 2) Spatially mapping where the magnetization dynamics are excited at particular field-frequency pairs of interest.  In Fig.~\ref{fig:tetra_optical} (c) and (d) we show the field-frequency spectrum of an individual surface and interior layer.  In (c) the spectroscopic weight of the lower frequency branch clearly lies on the surface layers, while in (d) the higher frequency branch is only present in the interior layers.  The spatial maps of the mode intensity on every layer for various field-frequency pairs are shown in (e).  In an external field of 500 Oe, the low-frequency mode occurs near 5.1 GHz, and the corresponding spatial maps show that the mode is nearly uniform and existing on the surface layers.  The high-frequency mode occurs at 8.9 GHz, and although there is some inhomogeneity in the spatial profile, it is a quasi-uniform excitation within the interior layers.  As the field is increased to 1000 Oe, the low frequency peak at 6.3 GHz still resides on the surface layers but the high frequency peak at 8.2 GHz has an appreciable amplitude on all layers.  Additionally, the mode across all layers is spatially inhomogeneous and tends to show localized effects on the surface.  At 1500 Oe, near the avoided energy level crossing, both the low and high frequency mode appear to be, spatially, more uniform.  The lower frequency mode at 7.1 GHz has a larger amplitude on the surface layers relative to the interior layers, and vice-versa for the high frequency mode at 7.6 GHz.  But, we emphasize that the non-vanishing amplitude on all layers for both eigenfrequencies provides clear evidence that the two optical modes, which are relegated to either surface or interior layers at low fields, are hybridized near 1500 Oe.  At 2000 Oe, both the low and high frequency modes are spatially inhomogeneous with a non-vanishing amplitude on all layers.

Based upon the behavior of the optical modes at low fields, as well as the spatial resolution of the modes, we have identified the low frequency branch as a surface layer dominant optical mode and the high frequency branch as an interior layer dominant optical mode.  This can be further confirmed by examining the phase difference of the $y$-components of the dynamic magnetization between individual layers.  In Fig.~\ref{fig:tetra_optical_phase}  (a) we plot the phase difference between a surface layer and interior layer, while in (b) we plot the phase difference between the two adjacent interior layers.  The maps are generated for the 5.1 GHz mode, at 500 Oe.  The phase difference between the surface and interior layers is more centered around 180$^\circ$, while the phase difference between the interior layers is centered around 0$^\circ$.  This implies that, although the surface layers are not directly exchange coupled, the dynamic components of the magnetization on the surface layers projected along the field ($y$-direction) are in phase.  This localized surface mode, is therefore an optical excitation.  In Fig.~\ref{fig:tetra_optical_phase} (c) and (d), we generate the same phase maps for the high frequency (8.9 GHz) mode at 500 Oe.  In the spatial regions where the dynamics are active on the surface, the phase difference between the surface and the interior mode tends to be centered around 0$^\circ$.  Between the interior layers, where this mode is mainly active, the phase difference is also centered around 0$^\circ$.  Thus, we have further verified that the high frequency mode localized to the interior layers is an optical excitation.  

\subsection{Tetralayer Acoustic Modes}

The field-frequency relationships for the acoustic modes of the tetralayer are shown in Fig.~\ref{fig:tetra_acoustic} (a) and (b) for the micromagnetic and macrospin model respectively.  Like the optical excitations, there is close agreement between both models.  Two acoustic branches are observed, and there is also an apparent avoided energy level crossing between each branch as the field is increased.  For readability, we have labeled the low-frequency branch as \textbf{I} and the high-frequency branch as \textbf{II} in (a) and (b).  In the low field limit, the low frequency branch has a linear dependence on the external field similar to what is calculated for the bilayer.  The new high frequency branch tends to decrease in frequency as the field is increased.  We immediately note that the zero-field frequency of this branch occurs at a lower frequency than the high frequency branch of the optical mode spectra seen in Fig.~\ref{fig:tetra_optical}.  Thus, even though the mode is excited with a field pulse which selects acoustic modes, the mode has traits which resemble an optical excitation.  Although this may appear to be a contradiction, this issue is resolved  within this section when mapping the phase difference in the magnetization dynamics between layers.  

In Fig.~\ref{fig:tetra_acoustic} (c) and (d) we examine the the acoustic mode spectra for the surface layers and the interior layers respectively.  It is clear that over a large field range, the lower frequency branch is excited on all layers.  At the highest fields, this branch does tend to localize preferentially on the interior layers.  The higher frequency branch is more prominent on the surface layers, but is also present on the interior layers.  At the largest fields, this mode tends to localize more on the surface layers.  In Fig.~\ref{fig:tetra_acoustic} (e), we spatially resolve the acoustic modes for select field-frequency pairs.  At 500 Oe, the low frequency (2.6 GHz) mode is spatially homogeneous and is almost equally excited across all layers; the field-frequency behavior combined with the uniform distribution of the mode across all layers suggests that this should be categorized as a uniform acoustic mode.  At the same field, the high frequency (7.0 GHz) mode is spatially inhomogeneous, with a non-vanishing amplitude across all four layers.  As the field is increased to 1000 Oe and 1500 Oe, the low frequency mode starts to localize on the interior layers and the higher frequency mode localizes on the surface layers.  The localization of the two modes is most readily apparent closest to the avoided crossing at 1500 Oe.  As the field is increased to 2000 Oe both the low and high frequency modes start to again to delocalize across all layers.  

\begin{figure}
\includegraphics[scale = .25]{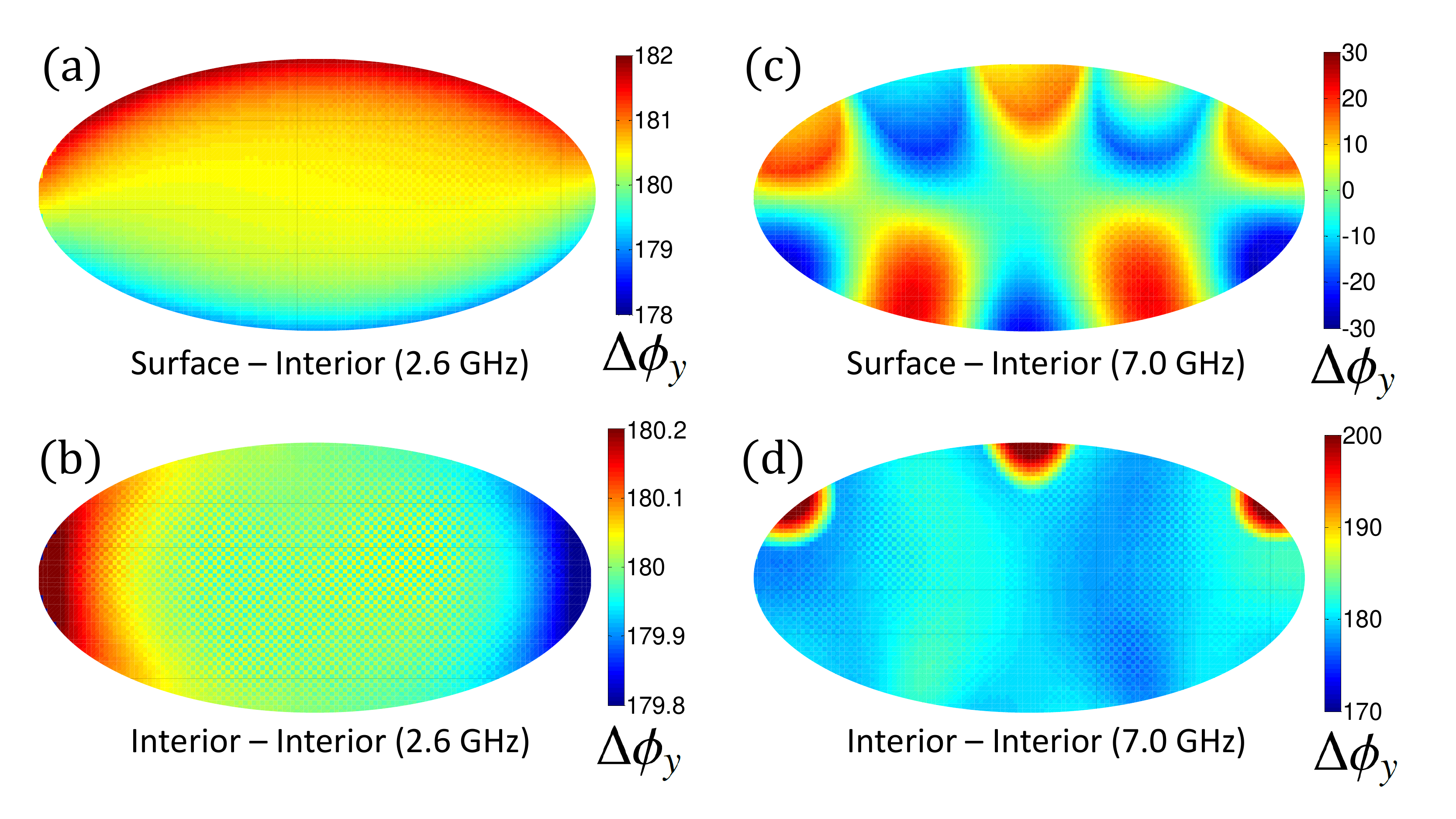}
\caption{(a) The phase difference, $\Delta \phi_y$, between a surface and interior layer of a tetralayer is spatially mapped at 2.6 GHz and 500 Oe.  (b) At the same field-frequency pair of (a), the phase difference between the two interior layers is spatially mapped.  Because the phase difference between all layers tends to be centered around $\Delta \phi_y$ = 180$^\circ$, this is identified as a uniform acoustic mode.  (c) The phase difference between a surface and interior layer of the mode at 7.0 GHz at 500 Oe is mapped.  (d)  For the same field-frequency pair in (c), the phase difference between the two interior layers is spatially mapped.  In (c), $\Delta \phi_y$ tends to be centered around 0$^\circ$ while in (d) $\Delta \phi_y$ is centered around 180$^\circ$.  This indicates that the higher frequency mode has a mixed optical-acoustic character depending upon what layer pairs are being considered.  }
\label{fig:tetra_acoustic_phase}
\end{figure}

The low frequency branch corresponds to a, uniform across-all-layers, acoustic mode.  This can be further confirmed by examining the phase difference of the $y$-components of the dynamic magnetization between individual layers.  In Fig.~\ref{fig:tetra_acoustic_phase}  (a) we plot the phase difference for between a surface layer and interior layer, while in (b) we plot the phase difference between the two adjacent interior layers.  The maps are generated for the 2.6 GHz mode, at 500 Oe.  Generally, both phase difference maps are spatially uniform with a phase angle centered around 180$^\circ$; this indicates an acoustic excitation.  In Fig.~\ref{fig:tetra_acoustic_phase} (c) and (d), we generate the same phase maps for the high frequency (7.0 GHz) mode at 500 Oe.  Importantly, the phase difference between the surface and the interior mode tends to be centered around 0$^\circ$, while the phase maps between the two interior layers is centered much closer to 180$^\circ$.  This implies that the interior layers have an acoustic-like behavior, while the interior-surface coupled dynamics is more like an optical excitation.  This mixed optical-acoustic character, is responsible for the zero-field and non-vanishing frequency near 7.0 GHz.    

In summary, when exciting the tetralayer with an external field pulse that selects acoustic magnons in the tetralayer geometry two modes are generated:  (1) A uniform acoustic mode across all layers and, (2) A quasi-uniform mode across all layers that has a mixed optical-acoustic behavior.  Increasing the external field forces the mode frequencies to approach one another and an avoided energy level crossing is observed.  Thus, the uniform acoustic mode, and the mixed optical-acoustic mode hybridize with one another.  A consequence of this hybridization is that the mode amplitudes localize on either the surface (high frequency branch) or the interior (low frequency branch) layers near the avoided crossing point.  This is in contrast to the optical modes, which start as localized excitations, only to become more uniform at higher fields near the avoided energy level crossing.  

\subsection{Hexlayer}
\begin{figure}
\includegraphics[scale = .38]{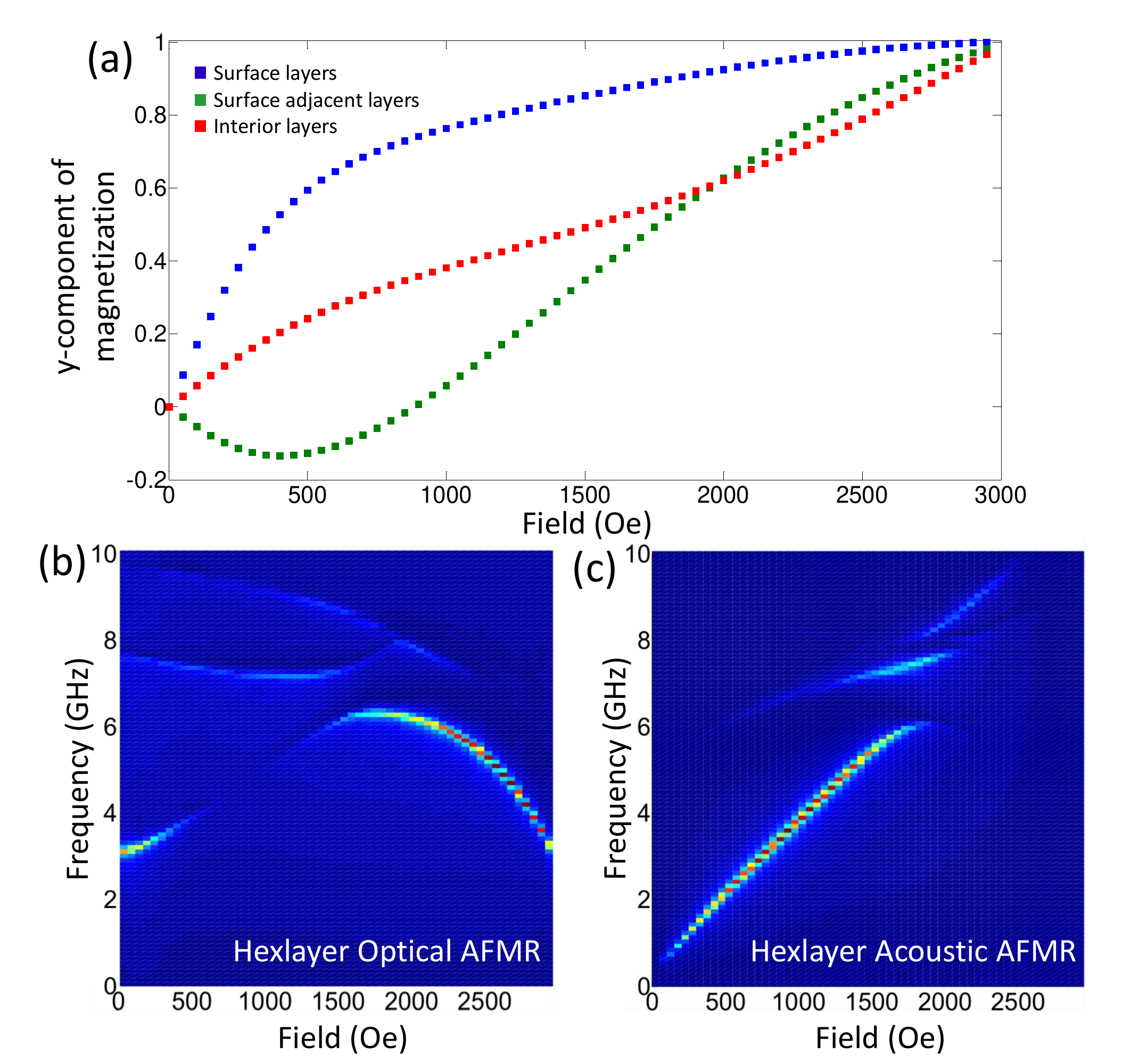}
\caption{(a) The $y$-component of the magnetization of the hexlayer is plotted on a layer-by-layer basis as a function of the external field.  Blue, green, and red squares correspond surface, surface-adjacent, and interior layers respectively.  (b) The optical AFMR spectra is plotted by taking an FFT of the $y$-component of the magnetization averaged across all six layers.  (c)  The acoustic  AFMR spectra is plotted by taking an FFT of the $z$-component of the magnetization averaged across all six layers.  For plots (b) and (c), the color scale on every individual plot is normalized to the largest amplitude in the field-frequency spectra.}
\label{fig:hexa}
\end{figure}

A macrospin analysis of the hexlayer (six layer system) is cumbersome, but employing micromagnetic simulations with additional layers provides no added technical difficulties. Here, we briefly examine both the optical and acoustic AFMR spectra of the hexlayer.  Qualitatively, moving from a tetralayer to a hexlayer adds two additional optical and acoustic AFMR magnon branches in the spectra.  This observation is explained by the fact that the surface layers, the layers adjacent to the surface, and the interior layers of the hexlayer \textit{all experience a different effective field} as a function of the external field.  In Fig.~\ref{fig:hexa} (a) this point is made manifest by looking at the $y$-component of the magnetization of individual layer pairs as a function of the external field.  There are three clearly different trajectories corresponding to the differences in how the magnetization in a given pair of layers rotates towards the applied field.  This rotation is non-trivial and somewhat counter-intuitive.  For example, in the low-field limit the layers that are adjacent to the surface layers \textit{rotate away from the external field}.  This can be understood by the fact that, in the low-field limit, the exchange interaction from the surface layers and the interior layers forces the surface-adjacent layers to rotate away from the field.  This counter-rotation allows the magnetization between layers to remain close to antiparallel even as a net magnetic moment is induced along the applied field direction.  The three different effective fields that each pair of layers feels at equilibrium is responsible for the three different magnon branches. 

In Fig. ~\ref{fig:hexa} (b) and (c) we plot the optical and acoustic AFMR spectra of the hexlayer.  Compared with the tetralayer the spectra is more complex.  Using the optical AFMR spectrum as an example, we see evidence for two avoided energy level crossings between all three optical AFMR branches.  This is due to the respective interactions of: 1) interior and surface-adjacent layer optical magnons, and 2) surface-adjacent and surface layer optical magnons.  If the tetralayer system illustrates how magnon modes in VdW magnets are sensitive to the layer number, the hexlayer system demonstrates that the complexity of the spectrum, in particular the number of avoided energy level crossings and hybridized magnonic excitations, also depends on the layer number.  

Although we do not show it here, we have verified that the octalayer (eight layer) has an additional optical and acoustic magnon branch relative to the hexlayer.  Thus, every time a pair of layers is added to the overall structure the magnon spectra changes by the addition of one optical and acoustic branch.  Throughout this process, the highest frequency optical branch does not tend to increase, and the additional magnon branches tend to be ``squeezed'' together towards the highest frequency branch.  This tendency may, in part, explain why the optical magnon modes that have been observed in bulk systems tend to have broader linewidths\cite{macneill2019gigahertz} i.e. the experimentally observed optical AFMR in bulk samples is a superposition of many closely spaced optical magnon modes.

\section{Engineering the Magnon Energy Spectra}
\begin{figure}
\includegraphics[scale = .18]{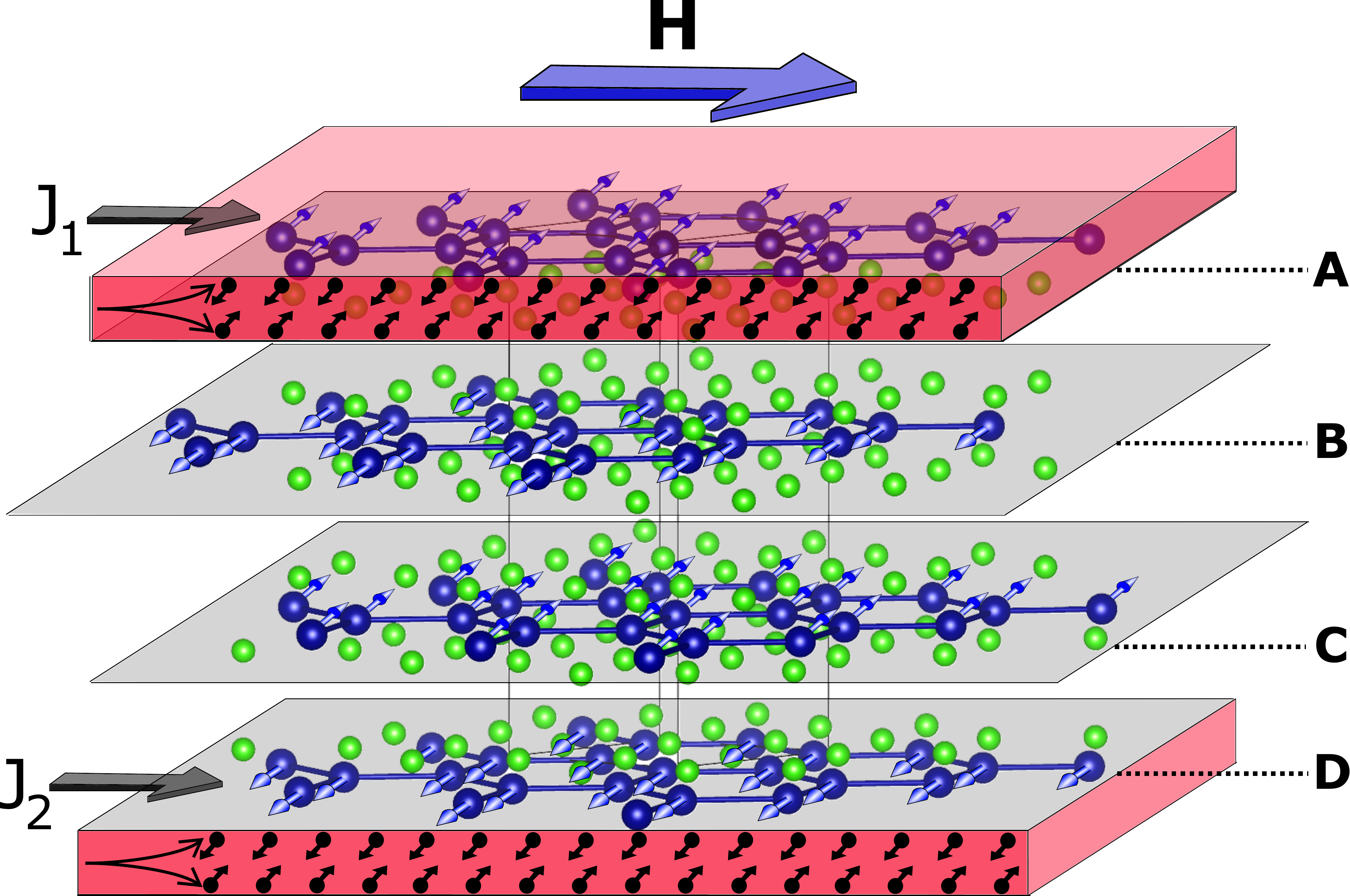}
\caption{A device form for manipulating the magnon-magnon interactions in VdW magnets consist of a CrCl$_3$ tetralayer that is encapsulated by two conducting layers. For this illustration, it is assumed that the conducting layers are spin Hall metals,however, Rashba-splitted and/or topological surface states are also potential candidates.  Thus, if two current densities, $J_1$ and $J_2$, pass through the conducting layers a spin accumulation develops on the surfaces.  The spin accumulation is illustrated with black circles and arrows that depict the polarization within the red layers.  As illustrated, if the spin polarization is parallel to the magnetization on the adjacent layers, ($\mathbf{A}$ and $\mathbf{D}$), a damping-like torque is exerted on the surface magnetization.}
\label{fig:magnon_dev}
\end{figure}

\begin{figure*}
\includegraphics[scale = .27]{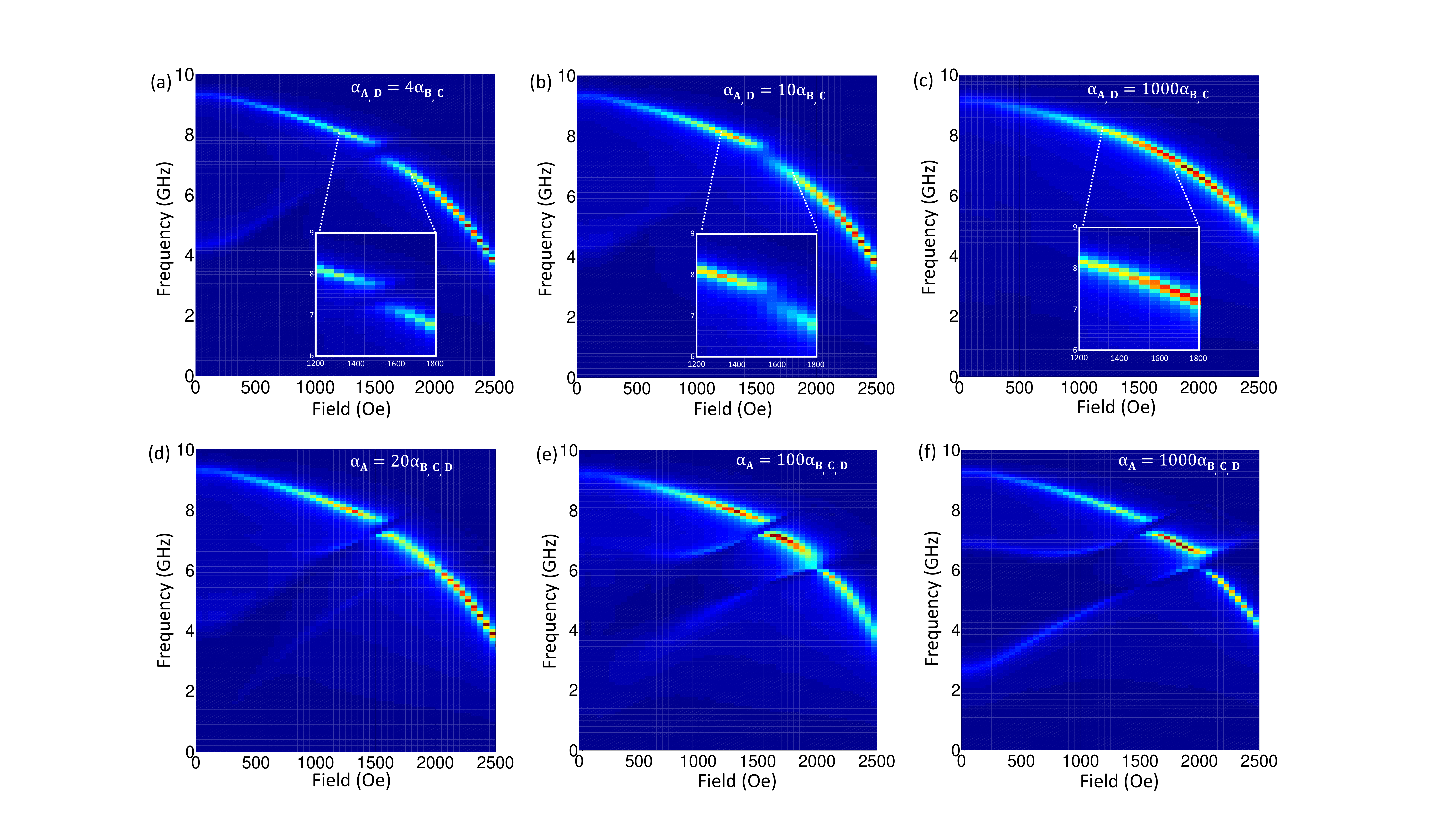}
\caption{In (a), (b), and (c) both surface layers experience increased damping relative to the interior layers.  The ratio of the damping on the surface to the interior is 4, 10, and 1000 for (a)-(c) respectively.  As the surface damping increases, the optical mode residing on the surface is suppressed and the avoided energy crossing vanishes.  In (d), (e), and (f) we asymmetrically modify the damping on only \textit{one} of the surface layers.  This asymmetric damping modification introduces a new magnon branch which can also interact with the primary optical magnon mode residing on the interior layers of the tetralayer.  Thus, two avoided energy level crossings appear in the magnon spectra.    }
\label{fig:magnon_eng}
\end{figure*} 

The previous section demonstrated a self-hybridization of both optical modes and acoustic modes in CrCl$_3$ for both tetralayers and hexlayers.  Self-hybridized magnons are unique because an obliquely-oriented external field is not needed to facilitate the magnon-magnon interaction.  The tetralayer is the simplest example which demonstrates how, in samples with more than two layers, multiple optical and acoustic modes are present. The dynamic interlayer exchange interaction is responsible for coupling these multiple modes together, and in this section we exploit this fact to demonstrate how the magnon spectrum can be manipulated by controlling the damping of the surface layers.

From a device perspective, the potential of tuning the interaction between the optical or acoustic modes without the use of an obliquely oriented field is attractive.   We consider one among many possible experimental architectures that can be used to electrically control the magnon-magnon interaction.  As illustrated in Fig.~\ref{fig:magnon_dev}, a CrCl$_3$ tetralayer is encapuslated within ``damping modifier'' layers that are colored red.  The damping modifier layers are intentionally generic.  Here, we intend them to be conducting materials which can pass two electrical current densities, $J_1$ and $J_2$.  If these conducting materials are spin Hall metals\cite{hoffmann2013spin}, such as Pt, Ta, or W,  they can be used to exert a damping- or antidamping-like torque on the surface layers.  The spin Hall effect of these metals would then generate a damping- or antidamping-like torque on the surface layers ($\mathbf{A}$ and $\mathbf{B}$) of the form $\boldsymbol{\tau}_{A(B)} \propto \pm \mathbf{m_{A(B)}} \times (\hat{y} \times \mathbf{m_{A(B)}})$\cite{sklenar2017unidirectional}.  This expression assumes a current passing along the $x$-direction of the damping modifiers, and the sign of the damping- or antidamping-like torque can be controlled by the current polarity.  It is also worth noting that one may select other van der Waals materials as damping modifiers as opposed to spin Hall metals.  Good candidates are Bi$_2$Se$_3$\cite{mellnik2014spin}, MoS$_2$\cite{zhang2016research}, and WTe$_2$\cite{macneill2017control}; all of which are cleavable materials known to generate damping-like torques.  

If both damping modifiers encapsulating the tetralayer are the same material, the application of a dc current through the entire stacked structure will either increase or decrease the damping in both layers simultaneously.  The utility of this proposed architecture is shown in Fig.~\ref{fig:magnon_eng}.  In this demonstration we limit ourselves to consider the optical magnon spectra in a tetralayer.  In (a), (b), and (c) the effective Gilbert damping parameter, $\alpha$, on the surface layers (\textbf{A} and \textbf{D}) is set to be 4, 10, and 1000 times greater than that of the interior layers.  As the damping of the optical magnon modes residing on the surface increases, the mode coupling that is mediated through the dynamic exchange interactions decreases.  Subsequently, the avoided energy level crossing, and the forbidden magnon frequencies vanish; only a single optical mode residing within the interior layers is present.

We also consider a more complex and equally interesting situation that occurs when it is assumed that the current is only passed through one of the two damping modifier layers in (d), (e), and (f).  Here, the effective Gilbert damping parameter on layer \textbf{A} is 20, 100, and 1000 times the damping of all other layers respectively.  It is seen that as the damping is increased for layer \textbf{A}, a third magnon branch which has a zero-field frequency just above 2 GHz emerges.  This new low frequency branch is a spatially uniform mode that is active on layers \textbf{B}, \textbf{C}, and \textbf{D}.  The phase difference $\Delta\phi_y$ between \textbf{B} $\&$ \textbf{C} suggests an acoustic excitation while the difference between \textbf{C} $\&$ \textbf{D} suggests an optical excitation.  This mixed character mode is not present in the original optical AFMR spectra for the tetralayer.  The intermediate frequency mode which is clearly seen in (e) and (f), is a surface mode that only resides on layer \textbf{D}.  The highest frequency branch is the original optical AFMR mode that resides within the interior layers.  In this configuration, the original avoided crossing remains because there is still a surface mode that is able to hybridize with the interior optical AFMR mode.  Interestingly, the new mixed character mode is also able to interact with the interior optical AFMR mode leading to a new avoided crossing in the energy spectra most clearly seen near a field of 2000 Oe near 6 GHz.     

In summary, we have demonstrated how to electrically  control  the magnon-magnon interactions in a tetralayer through the manipulation of the damping of the surface layers with spin-torques.  These manipulations can lead to both the creation and the elimination of avoided energy level crossings in the magnon spectra.  If both surface layers are damped, the optical avoided energy level crossing vanishes. If only one surface layer is damped, a new magnon branch appears and a second avoided energy level crossing is directly engineered into the overall spectrum. 

\section{Discussion and Conclusions}
Before concluding, a discussion of the implications that these results have for synthetic antiferromagnets is warranted. In synthetic AFMs, an interlayer exchange interaction between two ferromagnetic layers is mediated by the Ruderman-Kittel-Kasuya-Yosida (RKKY) interaction.  Depending on the interlayer thickness, the RKKY can then be used to facilitate a weak antiferromagnetic interaction similar to the interlayer exchange coupling in VdW magnets.  In synthetic antiferromagnets, ruthenium is commonly used as a spacer layer.  In stacks, where two ferromagnets are spaced with a Ru layer, the optical AFMR can often be found in a range of frequencies between 10-20 GHz \cite{li2016engineering, li2016tunable, waring2020zero}.  A good example of strong similarities between VdW antiferromagnets and synthetic antiferromagnets are strongly seen in a recent work where optical and acoustic magnons, in a stack of CoFeB/Ru/CoFeB\cite{sud2020tunable}, hybridize the same way that optical and acoustic magnons hybridized in bulk CrCl$_3$\cite{macneill2019gigahertz}.  Another recent result in synthetic antiferromagnets shows that hybridization between acoustic and optical magnon can be achieved without the use of an obliquely oriented field, provided that the magnons have a large enough wavevector\cite{shiota2020tunable}.  Clearly, strategies to facilitate magnon-magnon interactions in layered antiferromagnets are being actively explored in both synthetic and VdW magnets.  So far, these strategies can be employed for both material systems, and we emphasize that our work here is relevant to both types of magnets.  With that said, VdW materials may have implicit advantages.  For example, magic-angle graphene is a remarkable demonstration of how the electronic properties of graphene are extraordinarily sensitive to stacking\cite{cao2018unconventional}.  If magnetic anisotropy is found to be sensitive to stacking, VdW magnets will have tunability surpassing that of synthetic magnetic counterparts.

Synthetic magnets have also recently become theoretically\cite{lee2015macroscopic, galda2016parity, yu2020higher} and experimentally\cite{liu2019observation} interesting as meta-materials that possess parity-time symmetry breaking effects and exceptional points. Exceptional points can appear in the eigenvalue structure of coupled LLG equations, and if exploited, can be used to dramatically alter the frequency of magnetic resonances among other properties.  Reaching exceptional points in synthetic magnets can be achieved by adjusting either the damping or coupling between magnetic layers\cite{lee2015macroscopic,liu2019observation}.  These ideas are very much in the spirit of the results that we highlight in Fig. ~\ref{fig:magnon_eng}.  We suggest that van der Waals magnets may be a useful \textit{real} material to use in the exploration of exceptional points in magnets, along with synthetic magnets counterparts.

In summary, using an expanded macrospin model as well as detailed micromagnetic simulations, we have  calculated the optical and acoustic AFMR spectra for van der Waals magnets in the ultrathin limit.  We find that both the number of optical and acoustic magnon branches, as well as the mode frequencies, are very sensitive to the number of layers.  As an external magnetic field is applied, the various magnon branches can approach one another; in the case of tetralayers and hexlayers, both optical magnons and acoustic magnons can ``self-hybridize'' with one another.  For example, optical magnons residing on surface layers can hybridize with other optical magnons that reside on interior layers of a tetralayer.  This leads to characteristic avoided energy level crossings in the magnon mode spectra.  Using micromagnetic simulations we were also able to to demonstrate how electrical control of the damping on the surface layers can be used to control magnon-magnon interactions by altering both the number and strength of the avoided energy level crossings in the spectra.  Taken as a whole, these results demonstrate that VdW magnets should be strongly considered for future studies of coherent magnon-magnon effects in antiferromagnets.  We also emphasize that our results  can immediately be used to aid and interpret ongoing fundamental experimental inquiries into the antiferromagnetic resonance spectra of VdW magnets.

\begin{center}
\textbf{ACKNOWLEDGEMENTS} 
\end{center}

We would like to acknowledge Axel Hoffmann for illuminating concerning links to synthetic antiferromagnets. W.Z. was  supported by the U.S. National Science Foundation under Grant No. ECCS-1941426, and U.S. AFOSR under Grant No. FA9550-19-1-0254. J.S. acknowledges early computing resource support from the University of Illinois at Urbana-Champaign through the National Science Foundation MRSEC program under NSF award number DMR-1720633.  J.S. was also supported through a Wayne State University startup fund.  


\end{document}